\documentclass[twocolumn,prb,groupedaddress,amsmath,amssymb,floatfix]{revtex4}
\usepackage{graphicx}
\usepackage{amssymb}
\usepackage{dcolumn}
\usepackage{bm}
\begin{document}

\title{Self-diffusion and Cooperative Diffusion in Semidilute Polymer Solutions as measured by Fluorescence Correlation Spectroscopy
}
\author{Ute Zettl$^1$, Sebastian T.\ Hoffmann$^1$, Felix Koberling$^2$, Georg Krausch$^3$, J\"org Enderlein$^4$, Ludger Harnau$^5$, Matthias Ballauff$^1$}
\email{Matthias.Ballauff@uni-bayreuth.de}
\affiliation{$^1$Physikalische Chemie I, University of Bayreuth,D-95440 Bayreuth, Germany\\
\hspace*{2cm}$^2$Picoquant GmbH, D-12489 Berlin, Germany\hspace*{2cm}\\
\hspace*{2cm}$^3$University of Mainz, Germany\hspace*{2cm}\\
$^4$3.\ Institute of Physics, Georg August University, D-37077 G\"ottingen, Germany\\
$^5$Max-Planck-Institut f\"ur Metallforschung, Heisenbergstr. 3, D-70569 Stuttgart, Germany, 
\\and Institut f\"ur Theoretische und Angewandte Physik, Universit\"at Stuttgart, 
Pfaffenwaldring 57, D-70569 Stuttgart, Germany
}\date{\today}

\begin{abstract}
We present a comprehensive investigation of polymer diffusion in the semidilute 
regime by fluorescence correlation spectroscopy (FCS) and dynamic light scattering 
(DLS). Using single-labeled polystyrene chains, FCS leads to the self-diffusion 
coefficient while DLS gives the cooperative diffusion coefficient for exactly the 
same molecular weights and concentrations. Using  FCS we observe a new fast mode 
in the semidilute entangled concentration regime beyond the slower mode which is 
due to self-diffusion. Comparison of FCS data with data obtained by DLS on the same 
polymers shows that the second mode observed in FCS is identical to the cooperative 
diffusion coefficient measured with DLS. An in-depth analysis and a comparison with 
current theoretical models demonstrates that the new cooperative mode observed in 
FCS is due to the effective long-range interaction of the chains through 
the transient entanglement network. 
\end{abstract}
\maketitle

\section{Introduction}
Diffusion in polymer solutions is among the oldest subjects of polymer physics. 
\cite{DeGennes79,doi:86} In general, transport by diffusion can be characterized by two 
diffusion coefficients: the self-diffusion coefficient $D_s$ and the cooperative diffusion 
coefficient $D_c$. $D_s$ describes the motion of one molecule relative to the 
surrounding molecules due to thermal motions while $D_c$ describes the motion 
of a number of molecules in a density gradient. \cite{adam:77,Cosgrove83,Kanematsu05,Brown91,LeBon99} 
The obvious importance of diffusion in polymer physics has led to a rather large number 
of studies of $D_c$ by dynamic light scattering (DLS), 
\cite{adam:77,Cosgrove83,Pecora85,Brown91,LeBon99,Min03,Kanematsu05} while $D_s$ can be 
obtained by pulsed-field gradient nuclear magnetic resonance 
\cite{Cosgrove83,LeBon99,Kanematsu05} and label techniques like forced Rayleigh 
scattering \cite{Hervet79} or fluorescence correlation spectroscopy (FCS). 
\cite{Zettl04,Liu05,Zettl07} However, in many cases $D_s$ and $D_c$ could not be obtained 
for the same homopolymer using the same technique. 
Such measurements would be very interesting since a central problem in the dynamics 
of semidilute entangled polymer solutions is the quantitative understanding of the 
interplay of self-diffusion and cooperative diffusion. Very recently it has been 
found theoretically that the coupling of self- and cooperative motion due to 
topological constraints is also important for rather stiff macromolecules. 
\cite{Bier08}

At present, DLS is certainly among the most accurate methods to measure $D_c$ and there is a number of careful studies conducted on polymer solutions. In principle, FCS is the method of choice for studying diffusion of single macromolecules in a matrix of same molecular weight giving $D_s$ or in a solution of polymers of different molecular weight (tracer diffusion \cite{ribe:04,gian:07,Cher09}). In opposite to DLS, FCS requires chains labeled by a stable fluorescing molecule. Moreover, the number of labels per macromolecules should be constant to arrive at results that can be directly compared to theory. Given these problems, the use of FCS for measurements of $D_s$ on synthetic polymers has been scarce so far. \cite{Zettl04,Liu05,Zettl07,Grab08} Moreover, the full potential of this method has not yet fully been exploited yet since FCS should also allow one to obtain $D_c$. \cite{Ricka89, Scalettar89}

Recently, a well-defined polymeric model system has been presented and used for quantitative FCS-measurements in dilute solution: \cite{Zettl04,Zettl07} Nearly monodisperse polystyrene chains have been prepared by anionic polymerization and subsequently labeled by single fluorescent dye. Since the molecular weight of the different samples span a wide range, these polymers provide a nearly ideal model system for exploring the chain dynamics over a wide range of molecular weights and concentrations. Using these labeled chains, we recently presented an in-depth study of the experimental FCS set-up \cite{Zettl04} as well as of the dynamics in dilute solution. \cite{Zettl07}

Here we pursue these studies further by presenting an investigation of polymer diffusion in the semi-concentrated regime by FCS. In order to obtain accurate data of cooperative diffusion, these studies are combine with DLS-measurements on exactly the same molecular weights and concentrations. Thus, $D_c$ and $D_s$ can now be obtained from identical systems and directly be compared. In the course of these studies we found that a second cooperative mode becomes visible in the FCS-experiments if the concentration exceeds a given value. This surprising finding prompted us to conduct a full theoretical analysis of both the FCS- as well of the DLS-data throughout the entire time scale and range of concentrations available by these experiments. In doing so we extend the theoretical modeling beyond the usual scaling laws. The entire study is devoted to a comprehensive understanding of polymer dynamics in solution ranging from the dilute state up to the onset of glassy dynamics.

The paper is organized as follows: After the Section Experimental we first present the FCS-data together with the finding of the new cooperative diffusion. In the subsequent section a quantitative modeling of the data in terms of an analytical theory will be given. In the last section special attention will be paid to possible practical applications of these findings to the spinning of nanofibers. A Conclusion will wrap up the entire discussion.

\section{Experimental Section}
\subsection{Dye Labeled Polystyrene}
All experiments reported here were carried out with linear polystyrenes having a narrow 
molecular weight distribution. For details of the synthesis see ref \cite{Zettl04}. 
The molecular weight and polydispersity of the polymers are summarized in Table~\ref{PS}.
The solutions for the FCS experiments were prepared in toluene p.\ a.\ grade by blending 
a constant concentration of $10^{-8}\;\mathrm{M}$ Rhodamine~B labeled polystyrene with 
varying amounts of unlabeled polystyrene from the same synthesis batch. Each labeled 
polymer carries only one dye molecule at one of its ends. To verify our results, 
additional solutions were prepared with varying labeled polystyrene and a constant amount 
of unlabeled polystyrene. We have used preparative gel permeation 
chromatography to separate labeled polymer and free dye molecules. \cite{Zettl04,Zettl06}
Therefore, the resulting dye-labled polymer does not contain any measurable amount of 
free dye molecules.

%
%
\begin{table}[hb]
\begin{center}
\caption{\label{PS} Molecular weight $M_w$, polydispersity index PDI=$M_w/M_n$ and 
hydrodynamic radius $R_h$ at infinite dilution of the polystyrenes used in the 
present study. The second and the third virial coefficients $A_2$ and $A_3$, respectively,
have been calculated using scaling laws taken from the literature ($A_2$: 
ref \cite{Kniewske83} and $A_3$: ref \cite{Min03}). $c^+$ is the concentration at which 
the second diffusion time appears in the FCS measurements.}
\vspace*{0.2cm}
\begin{tabular}{cccccc}
\hline\\[-10pt]
$M_w[\frac{kg}{Mol}]$ & PDI & $R_h[nm]$ & $A_2[\frac{cm^3Mol}{g^2}]$ & 
$A_3[\frac{cm^6Mol}{g^3}]$ & $c^+[\mathrm{wt\%}]$ \\[-10pt]\\\hline\hline
11.5& 1.03& 1.(4)& 7.4$\cdot 10^{-4}$& 2.1$\cdot 10^{-3}$& -\\
17.3& 1.03& 1.(6)& 6.8$\cdot 10^{-4}$& 2.6$\cdot 10^{-3}$& -\\
67.0& 1.05& 3.(9)& 5.1$\cdot 10^{-4}$& 5.8$\cdot 10^{-3}$& 20\\
264 & 1.02& 7.(3)& 3.8$\cdot 10^{-4}$& 1.3$\cdot 10^{-2}$& 6.5\\
515 & 1.09& 9.(8)& 3.3$\cdot 10^{-4}$& 1.9$\cdot 10^{-2}$& 4.8\\
\hline
\end{tabular}
\end{center}
\end{table}
%
%

\subsection{Methods}
For FCS measurements we modified the commercial ConfoCor2 setup (Carl Zeiss, Jena, Germany) 
\cite{Rigler01} with a 40$\times$ Plan Neofluar objective (numerical aperture NA=0.9). 
The Rhodamine~B labeled PS-chains were excited by a HeNe-Ion laser at 543~nm. The intensity 
for all measurements was $4\mu$W in sample space. As second setup we used a MicroTime200 
(PicoQuant, Berlin, Germany) \cite{Wahl04} with a 100$\times$ oil immersion objective 
(NA=1.45). Here the detection beam path was divided by a 50/50 beam splitter on two 
detectors to crosscorrelate the signals. This crosscorrelation is necessary to prevent 
distortion of the fluorescence correlation function by detector afterpulsing. \cite{Enderlein05} 
For details of the FCS-measurements see refs \cite{Rigler01,Zettl04,Zettl07}. 

Cooperative diffusion coefficients 
$D_c$ were measured by DLS using an ALV 4000 light scattering goniometer (Peter, Germany).

\subsection{Evaluation of Data}
In FCS \cite{Magde72, Rigler01} a laser beam is focused by an objective with high 
numerical aperture (typically $\geq$~0.9) and excites fluorescent molecules entering the 
illuminated observation volume. The emitted fluorescent light is collected by the same 
optics and separated from scattered light by a dichroic mirror. The emitted light is 
detected by an avalanche photo diode. The time dependent intensity fluctuations $\delta 
I(\tau)=I(\tau)-\left\langle I(\tau) \right\rangle$ are analyzed by an autocorrelation function, 
where $\left\langle\,\,\, \right\rangle$ denotes an ensemble average. This autocorrelation 
function can be written as \cite{Ricka89}
\begin{equation} \label{eq1}
G(\tau)=\frac{1}{N}\int d {\bf q}\,\Omega ({\bf q}) C({\bf q},\tau)
\end{equation}
where \mbox{$\Omega ({\bf q})=\pi^{-\frac{3}{2}}w_{x,y}^2w_z 
\exp(\!-w_{x,y}^2 (q_x^2\!+\!q_y^2)/4\!-\!w_z^2 q_z^2/4)$}
is a Gaussian filter function characterizing the observation volume in Fourier space
with $\int d {\bf q}\, \Omega ({\bf q})=1$,  $N$ is the average number of fluorescently 
labeled molecules in the observation volume, and ${\bf q}=(q_x, q_y, q_z)$. Here 
$w_{x,y}=296$ nm is the dimension of the observation volume perpendicular to the 
optical axis and  $w_z=8 w_{x,y}$ is the dimension along the optical axis. 
\cite{Zettl04,Zettl07} For an ideal gas consisting of non-interacting molecules the 
initial amplitude reduces to the familiar relationship $G(0)=1/N$. \cite{Rigler01}

The time-dependent fluorescence density-density autocorrelation function 
$C({\bf q},\tau)$ is expressed in terms of a coupled-mode model \cite{Pusey82,Akcasu:91} 
as 
\begin{equation}  \label{eq2}
C({\bf q},\tau)=\frac{C_c(q,0)e^{-q^2 \phi_c(\tau)/6}+C_s(q,0)e^{-q^2 \phi_s(\tau)/6}}
{C_c(q,0)+C_s(q,0)}
\end{equation} 
where $q=|{\bf q}|$. Here the mean square displacements $\phi_c(\tau)$ and $\phi_s(\tau)$
are given by 
\begin{eqnarray}  \label{eq3}
\phi_c(\tau)&=&6 D_c \tau\,,
\\\phi_s(\tau)&=&6 D_s \tau+ B_s(\tau)\,. \label{eq4}
\end{eqnarray} 
The term $B_s(\tau)$ allows one to take into account the contributions from internal 
polymer chain motions. \cite{doi:86} If only a few of the molecules are fluorescently 
labeled, the self-diffusion coefficient $D_s$ can be measured in the FCS experiment.
\cite{Zettl07} If all of the molecules are fluorescently labeled, the cooperative diffusion
coefficient $D_c$ can be obtained. \cite{Scalettar89} In the case that neither of these 
limits applies, both the self mode and the cooperative mode will be present in the spectrum 
of the autocorrelation function. The diffusion coefficients can be extracted by fitting
\begin{eqnarray} 
\lefteqn{G(\tau)=}\nonumber
\\&&\sum_{i \in \{s,c\}} G_i(0) 
\left(1+ \frac{2 \phi_i(\tau)}{3 w_{x,y}^2}\right)^{-1}
\left(1+ \frac{2 \phi_i(\tau)}{3 w_{z}^2}\right)^{-1/2}
\nonumber
\\&&   \label{eq5}
\end{eqnarray} 
to the experimental data. FCS is not only sensitive to intensity fluctuations due 
to the motion of labled molecules but also due to photokinetic processes of the 
fluorescent dyes which occur for short times $\tau \lesssim 5\times 10^{-3}$ ms. 
This additional relaxation has been taken into account as discussed in refs 
\cite{Zettl04,Zettl06,Zettl07}.

DLS allows one to measure the time dependent autocorrelation function of the scattered 
electric field which can be expressed in terms of the elements of the fluid 
polarizability tensor. \cite{Pecora85} For an incident light wave traveling in the 
$x$ direction with a  polarization vector in the $z$ direction the intensity of the 
scattered electric field can be written as 
\begin{eqnarray}  
I_{VV}({\bf q},\tau)&\sim&\int d{\bf r}\,d{\bf r}'\,
\left\langle \alpha_{zz}({\bf r}+{\bf r}',\tau) \alpha_{zz}({\bf r}',0)\right\rangle
e^{i{\bf q}\cdot{\bf r}}\,,\nonumber
\\&&       \label{eq5a}
\end{eqnarray}
where the absolute value of the scattering vector ${\bf q}$ is given by 
$q=|{\bf q}|=(4\pi n/\lambda)\sin(\theta/2)$ in which $n$ is the refractive index 
of the medium. $\lambda$ is the incident wavelength and $\theta$ is the scattering 
angle. The $zz$ element of the fluid polarizability tensor is denoted as 
$\alpha_{zz}({\bf r},\tau)$. The experimentally accessible quantity is the intensity autocorrelation function $g^{(2)}_{VV}({\bf q},\tau)$. For photon counts 
obeying Gaussian statistics, the intensity autocorrelation function is related to 
the electric field autocorrelation function $g^{(1)}_{VV}({\bf q},\tau)$ according to
\begin{eqnarray} \label{eq5b}
g^{(2)}_{VV}({\bf q},\tau)&=&1+f_{VV} \left(g^{(1)}_{VV}({\bf q},\tau)\right)^2\,,
\end{eqnarray}
where $f_{VV}$ is dependent on the scattering geometry. The electric field correlation
function can be calculated for various systems. For a solution containing purely 
diffusing particles the electric field correlation function is given by 
$g^{(1)}_{VV}(q,\tau)=\exp(-q^2 D_c \tau)/\sqrt{f_{VV}}$.

\section{Diffusion coefficients measured by FCS}
%
%
\begin{figure}[hb]
\begin{center}
\includegraphics[width=7cm]{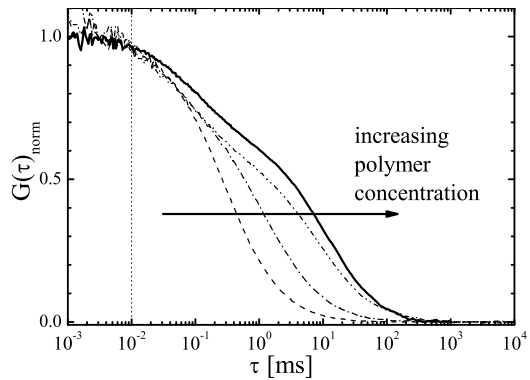}
\caption{Normalized autocorrelation function obtained from FCS for polystyrene of 
molecular weight $M_w=67~\mathrm{kg/Mol}$ in toluene for various 
polymer concentrations: 0.03~wt\% (-- --), 9.1~wt\% (-- $\cdot$), 20~wt\% (-- $\cdot\cdot$) 
and 28~wt\% (---). A second diffusion time appears at 20~wt\% on a shorter timescale compared 
to self-diffusion. The thick solid line is the normalized crosscorrelation curve without 
detector afterpulsing for the 28~wt\% polymer solution. The dotted vertical line marks the 
time scale above which this artefact becomes negligible, i.e., the solid thin and thick lines 
coincide for $\tau>0.01$ ms.}
\label{AC}
\end{center}
\end{figure}
%
%
Figure \ref{AC} shows normalized autocorrelation functions measured by FCS. The average number of 
labeled polymers in the observation volume was kept constant to $N\approx0.8$ whereas the number of unlabeled polymers 
increases up to $N_u=3 \times 10^6$ for the \mbox{28 wt~\%} polymer solution. The thin broken 
curves are measured at the ConfoCor2 setup and the thick solid curve is measured at the 
MicroTime200 setup. The curves obtained at the ConfoCor2 setup have an additional decay on 
the time scale less than 10~$\mu s$. This additional decay belongs to detector afterpulsing. 
Hence, the evaluation of the correlation curves has been done only for $\tau\ge$ 10~$\mu s$ 
as indicated by the dotted line in Figure \ref{AC}. For low polymer concentrations we obtained 
correlation curves with a single diffusion time. With increasing polymer concentration the
correlation curves shift to higher diffusion times. 

As an entirely new finding, Figure \ref{AC} presents a new mode related to a second diffusion time measured with 
FCS at higher polymer concentrations. This second diffusion time appears at shorter 
time scales than the one related to self-diffusion. The concentration $c^+$ at which the second
diffusion time is detected depends on the molecular weight: The higher the molecular weight, 
the lower is $c^+$ (see Table \ref{PS}). In general $c^+$ is about 15$\times$ the overlap
concentration determined in an earlier study. \cite{Zettl07} For the concentration $c^+$ 
the ratio between these two diffusion times is in the range of 60. From both diffusion times 
we calculated the diffusion coefficients from the relations given above. 

%
%
\begin{figure}[ht]
\begin{center}
\includegraphics[width=7cm]{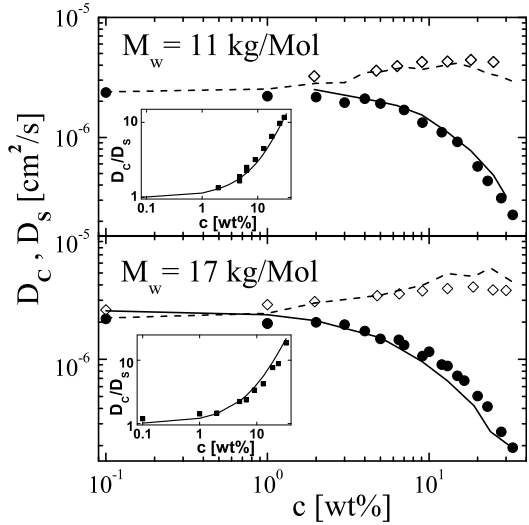}
\caption{Comparison of self-diffusion coefficients ($D_s$, $\bullet$) with cooperative 
diffusion coefficients ($D_c$, $\lozenge$) for different molecular weights: $M_w=$ 11 and 
17 kg/Mol (top and bottom). Open and solid symbols refer to DLS and FCS measurements, 
respectively. The solid lines represent $D_s$ calculated according to 
eq \ref{eq6} with $D_c$ as input from DLS measurements. The dashed lines represent $D_c$ 
calculated vice versa, i.e., with $D_s$ as input from FCS experiments. Insets: Measured 
ratio $D_c/D_s$ (symbols) together with the corresponding ratio obtained from
eqs \ref{eq6} and \ref{eq7} within a third order virial approximation (see Table \ref{PS}).}
\label{DSversusCshortMW}
\end{center}
\end{figure}
%
%
%
\begin{figure}[ht]
\begin{center}
\includegraphics[width=7cm]{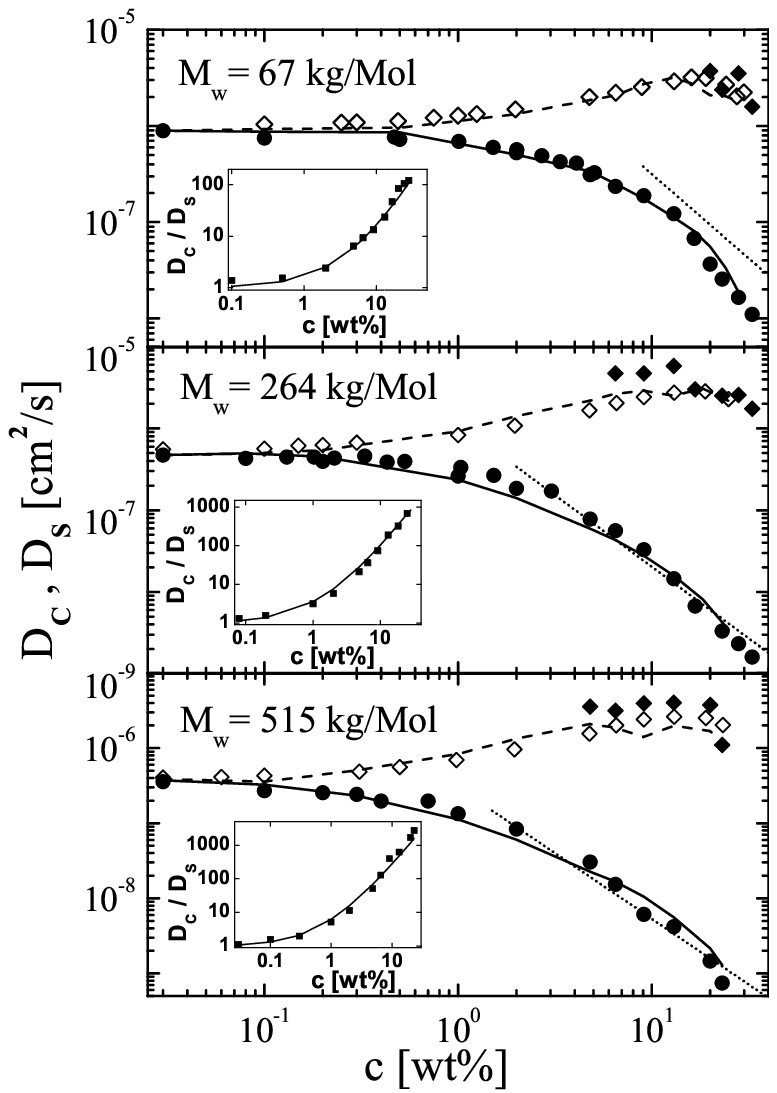}
\caption{Comparison of self-diffusion coefficients ($D_s$, $\bullet$) with 
cooperative diffusion coefficients ($D_c$, $\blacklozenge$, $\lozenge$) 
for different molecular weights: $M_w=$ 67, 264 and 515 kg/Mol (from top to bottom). Open 
and solid symbols refer to DLS and FCS measurements, respectively. The solid lines represent 
$D_s$ calculated according to eq \ref{eq6} with $D_c$ as input from DLS measurements. 
The dashed lines represent $D_c$ calculated vice versa, i.e., with $D_s$ as input from 
FCS experiments. For comparison the dotted lines represents the scaling prediction 
$D_s \sim M_w^{-2} c^{-7/4}$ for long polymer chains in the semidilute entangled regime 
(see eq \ref{eq11}). Insets: Measured ratio $D_c/D_s$ (symbols) together with the 
corresponding ratio obtained from eqs \ref{eq6} and \ref{eq7} within a third order virial 
approximation (see Table \ref{PS}).}
\label{DSversusClargeMW}
\end{center}
\end{figure}
%
%

In Figures \ref{DSversusCshortMW} and \ref{DSversusClargeMW} all diffusion 
coefficients measured with FCS and DLS are compared at identical conditions. 
At infinite dilution both diffusion coefficients $D_s$ and $D_c$ have the
same value. In dilute solutions $D_s$ and $D_c$ show a linear dependency on the concentration 
as expected according to the Kirkwood-Riseman theory. \cite{Kirkwood48} But $D_s$ decreases
whereas $D_c$ increases with increasing polymer concentration. The decrease of $D_s$ is due 
to the friction between the chains and the increase of $D_c$ is due to the increasing osmotic
pressure. \cite{Fujita90,Kim86} At high concentrations $D_c$ exhibits a maximum.

The insets in Figure \ref{DSversusCshortMW} and Figure \ref{DSversusClargeMW} show the ratio ${D_c}/{D_s}$ of measured values. The lines are theoretical values calculated according to \cite{Kanematsu05,LeBon99} 
\begin{equation}  \label{eq6}
\frac{D_c}{D_s} = \left(1-\bar{v} c\right) \frac{d \Pi}{d c}
\end{equation}
with the partial specific volume of the polymer $\bar{v}$ and the polymer concentration 
$c$. The dependence of the osmotic pressure on $c$ can approximated by a virial expansion
\begin{equation}  \label{eq7}
\frac{d \Pi}{d c}= 1 + 2A_2 M_w c + 3A_3 M_w c^2 + \ldots\, ,
\end{equation}
where $A_2$ and $A_3$ are the second and third virial coefficients, respectively, and $M_w$ 
is the molecular weight. For the calculation of $d \Pi\,/dc$ we used the corresponding values 
from the literature gathered in Table \ref{PS} and \mbox{$\bar{v}=0.916$ cm$^3$/g}. \cite{Schulz57} 
The measured and the calculated ratio are well described as demonstrated by the inset of 
Figures \ref{DSversusCshortMW} and \ref{DSversusClargeMW}. The self-diffusion coefficients
$D_s$ can be determined from the cooperative diffusion coefficient $D_c$ obtained by DLS
measurements and vice versa. $D_s$ and $D_c$ can be measured with high accuracy by FCS 
and DLS using the same polymers. Their relation is fully understood in terms of eq \ref{eq6}. 
For comparison we note that both the molecular dye diffusion coefficient and
the macromolecular tracer diffusion coefficient decrease with increasing 
concentration of the matrix polymer. \cite{Cher09}

%
\begin{figure}[ht]
\begin{center}
\includegraphics[width=7cm]{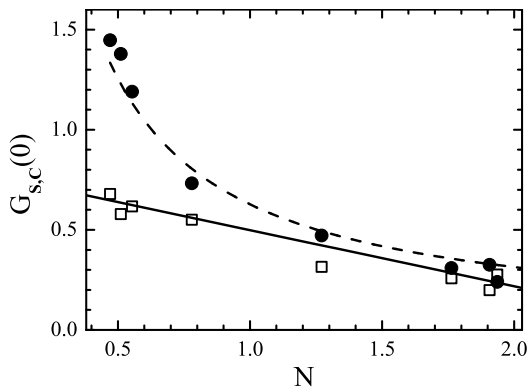}
\caption{Amplitudes $G_s(0)$ ($\bullet$) and $G_c(0)$ ($\square$) 
extrapolated from the measured FCS-autocorrelation function $G(\tau)$ as a 
function of labeled molecules $N$ for polystyrene with $M_w=67~\mathrm{kg/Mol}$ at 20~wt\%. 
For the self-diffusion $G_s(0)\propto 1/N$ (-- --), while $G_c(0)$ exhibits a linear dependence 
on $N$ (---) for the cooperative diffusion.}
\label{fig4} 
\end{center}
\end{figure}
%
%

Figure \ref{fig4} displays the amplitudes $G_i(0)$ (see eq \ref{eq5}) as a function 
of $N$ for polystyrene with $M_w=67~\mathrm{kg/Mol}$ at 20~wt\%. The amplitude of the 
self-diffusion mode $G_s(0)$ is proportional to $1/N$. In the presence of non-correlated 
background signal (scattering, afterpulsing, electronic noise) this is modified to $1/N - 
2b/N^2$. \cite{Rigler01}  Here $b$ is proportional to the noise 
intensity, which is assumed to be significantly smaller than the fluorescence signal. 
For the cooperative mode one finds an amplitude scaling of $1 - 2bN$.
For sufficiently small $b$, this will yield a dependence as shown by Figure \ref{fig4} for 
the fast correlation component.

The ratio $G_c(0)/G(0)$ is a non-monotonic function of the concentration for a fixed 
number of labeled molecules $N$. It increases form 0 to a value below 1 at the concentration 
$c^+$. $G_c(0)/G(0)$ slightly decreases upon further increasing the concentration in 
the semidilute entangled regime. Finally, it increases upon approaching the glass 
transition concentration.

\section{Scaling theory and Langevin equation approach}
In the following section, the findings presented in the previous sections will be compared 
to current models of polymer diffusion.

\subsection{Scaling theory and reptation model}
The application of scaling theory and the reptation model to polymer solutions has 
been presented in various treatises (see, e.g., refs \cite{DeGennes79,doi:86,lodg:90,mcle:02}). 
Hence, we only discuss the equations necessary for this study. Three concentration regimes 
can be distinguished: dilute, semidilute unentangled, and semidilute entangled solutions.
Scaling arguments and the reptation model lead to following relations for the self-diffusion 
coefficient $D_s$ and the cooperative diffusion coefficient $D_c$:
\begin{eqnarray} 
D_s &=& D_c \sim M_w^{-3/5}\, c^{0} \,,\hspace*{0.5cm} c \ll c^*\,, 
\label{eq9}
\\D_c &\sim & M_w^0\, c^{3/4}      \,,\hspace*{0.5cm} c > c^*\,, 
\label{eq10}
\\D_s &\sim & M_w^{-2}\, c^{-7/4}  \,,\hspace*{0.5cm} c > c^{**}\,.
\label{eq11}
\end{eqnarray}
Here the overlap concentration $c^*$ is the boundary concentration between the dilute 
and semidilute regimes. This concentration depends on molecular weight as
\begin{equation} \label{eq8}
c^* \sim M_w^{1-3\nu}=M_w^{-4/5}\,,
\end{equation}
where the Flory exponent $\nu=3/5$ for a good solvent has been used. The crossover 
concentration from the semidilute unentangled to the semidilute entangled regime is 
denoted as $c^{**}$.

\begin{figure}[ht]
\begin{center}
\includegraphics[clip=,width=7cm]{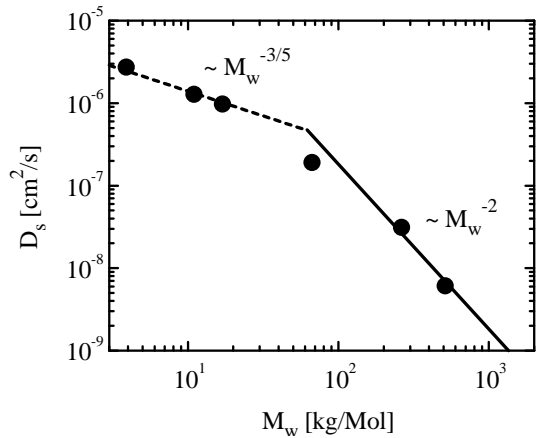}
\caption{The self-diffusion coefficient $D_s$ ($\bullet$) measured by FCS at the fixed 
concentration \mbox{$c=9.1$ wt \%} as a function of the molecular weight $M_w$.
The dashed and solid lines of slope $M_w^{-3/5}$ (see eq \ref{eq9}) and 
$M_w^{-2}$ (see eq \ref{eq11}), respectively, represent two asymptotic scaling 
regimes.}
\label{fig5}
\end{center}
\end{figure}

For very low concentrations in the dilute regime, the self-diffusion coefficient 
is indistinguishable from the cooperative diffusion coefficient as is apparent 
from Figures \ref{DSversusCshortMW} and \ref{DSversusClargeMW}. 
In Figure \ref{fig5} the self-diffusion coefficient is plotted as a function of 
the molecular weight $M_w$ for a fixed concentration \mbox{$c=9.1$ wt \%}. The 
experimental data (solid squares) follow the scaling laws given by eq \ref{eq9}
(dashed line) and eq \ref{eq11} (solid line) for \mbox{$M_w \le 20$ kg/Mol} and 
\mbox{$M_w \ge 264$ kg/Mol}, respectively. Moreover, $D_s$ is rather independent 
of concentration for \mbox{$c \lesssim 10$ wt \%} in the case of the low 
molecular weight solution (see Figure \ref{DSversusCshortMW} and eq \ref{eq9}).
The concentration dependence of $D_s$ of the higher molecular weight solutions 
(\mbox{$M_w \ge 264$ kg/Mol}) is in accord with the scaling prediction for the 
reptation model (eq \ref{eq11}) which is represented in Figure \ref{DSversusClargeMW} 
by the dotted lines. Hence the FCS measurements verify the basic scaling and reptation 
theory for semidilute entangled polymer solutions similar to earlier forced 
Rayleigh scattering experiments of polystyrene in benzene. \cite{Hervet79,lege:81}
%
%
%
%

In the limit $c \to 0$ the experimental data follow the scaling law given by 
eq \ref{eq9} irrespective of the molecular weight, \cite{Zettl07} i.e., 
also the higher molecular weight PS solutions obey the scaling relation
$D_s \sim M_w^{-3/5}\, c^{0}$ . This result is in agreement with earlier 
quasi-elastic light scattering experiments for polystyrene in 2-butanone 
\cite{king:73} or in benzene. \cite{adam:77}

\subsection{Internal motions of chains}
In order to examine the influence of internal chain motions such as bending 
and stretching on the dynamics (see refs \cite{harn:95,harn:98,harn:99a,harn:99b} 
and references therein), one may trace out the internal degrees 
of freedom of a polymer chain by studying the monomer mean square displacement 
$B_s(\tau)$ in eq \ref{eq4} in detail. Various theoretical predictions 
on the time dependence of the monomer mean square displacement 
of both continuously and single labeled DNA molecules in aqueous solution have been 
verified using FCS measurements. \cite{lumm:03,shus:04,gros:06,petr:06,wink:06,shus:08}
In these earlier experimental and theoretical studies the $\Theta$ condition 
has been considered. However, for PS in toluene solutions the intramolecular excluded 
volume interaction has to be taken into account. In this case scaling arguments 
\cite{krem:84,paul:91} lead to the following time dependence of the monomer mean 
square displacement:
\begin{eqnarray} 
B_s(\tau) = B_s \tau^{1/(1+1/(2\nu))} = B_s \tau^{6/11}\,.
\label{eq12}
\end{eqnarray}
It proves convenient to consider the function \mbox{$1/G(\tau)-1$}, which amplifies 
the time dependence of $G(\tau)$ for small times, because $w_z^2=64 w_{x,y}^2$ 
in eq \ref{eq5}. \cite{wink:06} If the autocorrelation function $G(\tau)$ exhibits a 
time dependence according to eqs \ref{eq4}, \ref{eq5}, and \ref{eq12} with $G_c(0)=0$, 
a double logarithmic plot will directly yield the exponent $1/(1+1/(2\nu))$ for small 
times provided the intramolecular dynamics dominates, i.e., $B_s(\tau) >> 6 D_s \tau$. 

\begin{figure}[ht!]
\begin{center}
\includegraphics[clip=,width=7cm]{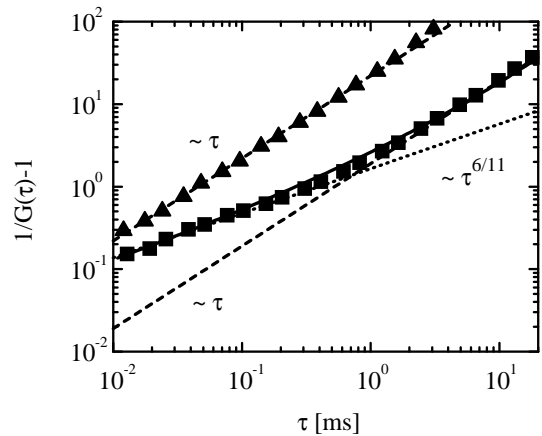}
\caption{The autocorrelation function $1/G(\tau)-1$ of a 515 kg/Mol ($\blacksquare$)
and a  17 kg/Mol ($\blacktriangle$) polystyrene solution measured by FCS in the limit 
\mbox{$c \to 0$} as a function of the time $\tau$. The dotted and dashed lines of 
slope $\tau^{6/11}$ (see eq \ref{eq12}) and $\tau$ (see eq \ref{eq4}), respectively, 
represent two asymptotic scaling regimes. The solid line displays the result 
for the 515 kg/Mol polystyrene solution as obtained from  eq \ref{eq5} with 
eqs \ref{eq4} and \ref{eq12} as input. The autocorrelation function of the 17 kg/Mol 
polystyrene solution ($\blacktriangle$ and upper dashed line) is shifted up by a factor 
of 2.}
\label{fig6}
\end{center}
\end{figure}

Figure \ref{fig6} shows such a representation of the autocorrelation function for 
the 515 kg/Mol PS chains in dilute solution. The experimental data (solid squares) follow 
the scaling law given by eq \ref{eq12} (dotted line) and the diffusive behavior 
(lower dashed line) for short and large times, respectively. Hence for short times 
the decay of the autocorrelation function is dominated by intramolecular chain relaxations, 
while self-diffusion dominates for large times. Figure \ref{fig6} demonstrates 
that the measured autocorrelation function agrees with the calculated 
results (solid line) obtained from eqs \ref{eq4}, \ref{eq5}, and \ref{eq12} 
with $D_s$ and $B_s$ as input. The mean displacements $\sqrt{\phi_s(\tau)}$ as calculated 
from eq \ref{eq5} with $G_s(0)=1$ and $G_c(0)=0$ are given by 131 nm and 598 nm
for $\tau=0.01$ ms and $\tau=1$ ms, respectively.

It is apparent from Figure \ref{fig6} that the contribution of internal chain motions 
cannot be observed in the case of  the 17 kg/Mol PS chains in dilute solution (solid 
triangles) because of the dominating diffusive motion (upper dashed line). The 
self-diffusion coefficient $D_s$ increases upon decreasing molecular weight according to 
eq \ref{eq9}, while $B_s$ is less dependent on molecular weight. Finally, it is 
worthwhile to mention the contribution of internal chain motions to the dynamics decreases
upon increasing the polymer concentration because of the presence of the surrounding 
polymer chains. \cite{krem:84,harn:99c} 
%
%
%
%

\subsection{Cooperative diffusion}
\begin{figure}[ht!]
\vspace*{0.5cm}
\begin{center}
\includegraphics[clip=,width=7cm]{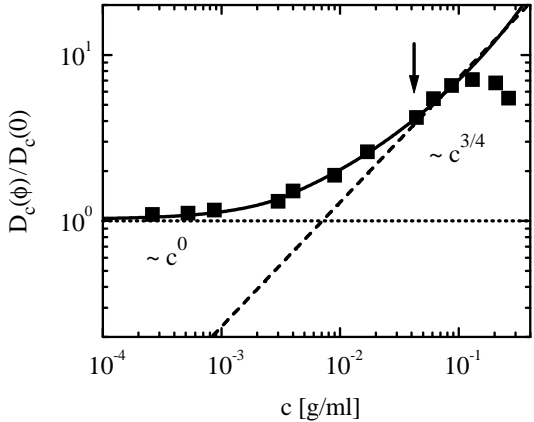}
\caption{The normalized cooperative diffusion coefficient $D_c$ ($\blacksquare$; DLS) 
of a 515 kg/Mol polystyrene solution as a function of the concentration $c$.
The dashed and dotted lines of slope $c^{3/4}$ (see eq \ref{eq10}) and 
$c^0$ (see eq \ref{eq9}), respectively, represent two asymptotic scaling 
regimes. The solid line displays  the results as obtained from the Langevin 
and generalized Ornstein-Zernike equation according to eqs \ref{eq15} - 
\ref{eq21}. The arrow marks the location of the concentration 
\mbox{$c^+=0.044$ g/ml} at which the cooperative diffusion mode appears in 
the FCS measurements (see Figure \ref{DSversusClargeMW}).}
\label{fig7}
\end{center}
\end{figure}

We now turn our attention to the scaling law for the cooperative diffusion coefficient
given by eq \ref{eq10}. Figure \ref{fig7} displays the cooperative diffusion 
coefficient $D_c$ of the 515 kg/Mol PS solution (solid squares) together with the 
scaling law (dashed line) as a function of the concentration. Several experimental 
measurements have yielded the concentration dependence $D_c \sim c^{0.65}$ instead 
of the scaling prediction $D_c \sim c^{3/4} = c^{0.75}$. 
\cite{adam:77,wilt:84,nemo:84,zhan:99,rauc:03} Various possible explanations for 
these deviations from the scaling law have been discussed, \cite{geis:78,brow:90}
such as the countermotion of the solvent induced by the motion of the polymers. 
On the basis of our results shown in Figure \ref{fig7} we note that the transition
between the dilute regime with $D_c \sim c^0$ (dotted line and eq \ref{eq9}) and 
the semidilute unentangled regime with $D_c \sim c^{3/4}$ (dashed line and 
eq \ref{eq10}) is not so abrupt, as has been assumed by scaling theories, but is
a rather smooth crossover that extends over more than one order in magnitude of 
concentration. 

It has been emphasized that it would be desirable to model 
the dynamics both in the dilute regime and the semidilute regimes explicitly within 
one theoretical approach. \cite{lodg:90} Successful models should incorporate 
the transition region between the dilute regime and the semidilute regimes.
In the next subsection we provide a quantitative basis for such a modelling of 
cooperative dynamical properties of polymer chains in good solution.
%
%
%
%

\subsection{Analytical theory: Langevin and generalized Ornstein-Zernike equation}
We consider a monodisperse polymer solution consisting of $N_{tot}=N+N_u$ polymer
chains and the solvent. Each polymer chain carries $n$ scattering units.
The total dynamic scattering function $S_{tot}(q,\phi,\tau)$ is defined as 
\begin{eqnarray}   \label{eq13}
S_{tot}(q,\phi,\tau)&=&\frac{1}{N_{tot} \, n^2}\left\langle
\sum\limits_{\alpha, \gamma=1}^{N_{tot}}
\sum\limits_{j,k=1}^{n}
e^{i{\bf q}\cdot \left({\bf r}_{\alpha j}(\tau)-{\bf r}_{\gamma k}(0)\right)}
\right\rangle\,,\nonumber
\\&&
\end{eqnarray}
where $q=|{\bf q}|$ is the magnitude of the scattering vector ${\bf q}$ and 
$\langle \,\,\,\, \rangle_\phi$ denotes an ensemble average for a given polymer 
volume fraction $\phi$. Here  ${\bf r}_{\alpha j}(\tau)$  is the position vector 
of the $j$-th scattering unit ($1 \le j \le n$) of the $\alpha$-th 
particle ($1 \le \alpha \le N_{tot}$) at time $\tau$. The normalized total 
dynamic scattering function is related to the electric field autocorrelation 
function measured by DLS according to 
$S_{tot}(q,\phi,\tau)/S_{tot}(q,\phi,0)=g^{(1)}_{VV}(q,\tau) \sqrt{f_{VV}}$. (see eq \ref{eq5b}).
The time evolution of the total dynamic scattering function is assumed to be governed 
by the Langevin equation \cite{doi:86}
\begin{equation}    \label{eq14}
\frac{d}{d\tau} S_{tot}(q,\phi,\tau)=-\Gamma(q,\phi) S_{tot}(q,\phi,\tau)\,.
\end{equation}
The validity of this equation is not obvious since entanglements have 
not been taken into account in the derivation of this equation. \cite{doi:86}
However, the short time-scale dynamics can be described by eq \ref{eq14}
since the topological constraints are not so important in the short time-scale 
dynamics as is apparent from the fact that the cooperative diffusion 
coefficient $D_c$ is considerably larger than the self-diffusion coefficient 
$D_s$ in the semidilute entangled regime (see Figure \ref{DSversusClargeMW}).
The decay rate $\Gamma(q,\phi)$ is given by \cite{doi:86}
\begin{eqnarray} 
\lefteqn{\Gamma(q,\phi)=}\nonumber
\\&&\frac{k_BT}{4\pi^2 \eta}\int\limits_0^\infty
d q_1\, q_1^2 \frac{S_{tot}(q_1,\phi,0)}{S_{tot}(q,\phi,0)}
\left(\frac{q_1^2+q^2}{2q_1q}
\log\left|\frac{q_1+q}{q_1-q}\right|-1\right)\,,\nonumber
\\&&          \label{eq15}
\end{eqnarray}  
where the temperature $T$ and the viscosity $\eta$ characterize the solvent. 
The volume fraction-dependent cooperative diffusion coefficient $D_c(\phi)$ can 
be calculated according to 
\begin{equation} \label{eq16}
D_c(\phi)=\lim_{q\to 0}\frac{\Gamma(q,\phi)}{q^2}\,.
\end{equation}
Furthermore, the total static scattering function reads 
\begin{equation} \label{eq17}
S_{tot}(q,\phi,0)=1+\phi h(q,\phi)/(V_p P(q,\phi))\,,
\end{equation}
where $V_p$ is the volume of a dissolved polymer chain and $h(q,\phi)$ 
is a particle-averaged total correlation function. The particle-averaged 
intramolecular correlation function
\begin{eqnarray}  
P(q,\phi)&=&\frac{1}{N_{tot} \, n^2}\left\langle
\sum\limits_{\alpha=1}^{N_{tot}}
\sum\limits_{j,k=1}^{n}
e^{i{\bf q}\cdot \left({\bf r}_{\alpha j}(0)-{\bf r}_{\alpha k}(0)\right)}
\right\rangle\,,\nonumber
\\&&    \label{eq18}
\end{eqnarray} 
characterizes the geometric shape of the polymer chains at a given volume
fraction $\phi$. While the particle-averaged intramolecular correlation function 
accounts for the interference of radiation scattered from different parts of the same 
polymer chain in a scattering experiment, the local order in the fluid is characterized by 
$h(q,\phi)$. The particle-averaged total correlation function is related to a 
particle-averaged direct correlation function $c(q,\phi)$ by the generalized 
Ornstein-Zernike equation of the Polymer Reference Interaction Site Model (PRISM), 
which reads (see refs \cite{schw:97,harn:08,yeth:09} and references therein) 
\begin{equation} \label{eq19}
h(q,\phi)=P^2(q,\phi)c(q,\phi)/(1-\phi c(q,\phi)P(q,\phi)/V_p)\,.
\end{equation}
This generalized Ornstein-Zernike equation must be supplemented by a closure 
relation. If the interaction sites are simply the centers of exclusion spheres, 
to account for steric effects, a convenient closure is the Percus-Yevick 
approximation. \cite{schw:97} The PRISM integral equation theory has been 
successfully applied to various experimental systems such as polymers,
\cite{schw:97,harn:01a}  bottle-brush polymers, \cite{boli:07,boli:09}
rigid dendrimers, \cite{rose:06,harn:07} and charged colloids.
\cite{yeth:96,yeth:97,shew:98,harn:00,harn:01,harn:02,webe:07,henz:08}

The overall size of the polymer chains is reduced considerably upon increasing 
the volume fraction implying a concentration dependence of the particle-averaged 
intramolecular correlation function $P(q,\phi)$. Therefore, we consider 
the following particle-averaged intramolecular correlation function \cite{fuch:97}
\begin{equation} \label{eq20}
P(q,\phi)=\left(1+0.549\, q^2 r_g^2(\phi)\right)^{-5/6}
\end{equation}
with the volume fraction dependent radius of gyration 
\begin{eqnarray}  \label{eq21}
r_g^2(\phi)&=& 
\left\{\begin{array}{c@{\quad,\quad}l}
r_g^2(0) & c < c^*
\\r_g^2(0)\left(\frac{\displaystyle c}{\displaystyle c^*}\right)^{-1/8} 
&c > c^*
\end{array}\right.\,.
\end{eqnarray}
Here the relation between the volume fraction $\phi$ and the concentration $c$ 
is given by $\phi= \bar{v} c$, where \mbox{$\bar{v}=0.916$ cm$^3$/g} is the 
specific weight of PS. \cite{Schulz57} The scaling law given by eq \ref{eq21}
has been confirmed experimentally for PS in a good solvent using small angle 
neutron scattering. \cite{daou:75} 

Figure \ref{fig7} demonstrates that the measured cooperative diffusion coefficients 
(solid squares) agree with the calculated results (solid line) obtained from 
eqs \ref{eq15} - \ref{eq21} both in the dilute and semidilute regimes. In particular, 
the features of the broad crossover region between the dilute and the semidilute 
regimes are captured correctly by the integral equation theory. In the calculations 
the model parameter \mbox{$c^*=0.0032$ g/ml$\,$} \cite{Zettl07} and 
\mbox{$r_g(0)=32.8$ nm} for the 515 kg/Mol PS solution has been used. This radius of 
gyration is about \mbox{6 \%} larger than corresponding radii of gyration of PS 
in various good solvents. \cite{Kniewske83,fett:94,Min03,tera:04} The deviation between 
the radius of gyration used in the calculations and the radii of gyration 
reported in the literature might be due to the fact that the hydrodynamic interaction 
has been taken into account in terms of the Oseen tensor in order to derive
eq \ref{eq15}. Using the Rotne-Prager tensor \cite{rotn:69,harn:96} 
as a first correction to the Oseen tensor will improve the results. Moreover, the 
size polydispersity $M_w/M_n=1.09$ of the 515 kg/Mol PS solution leads to a diffusion 
coefficient which is characteristic for monodisperse polymers of larger radius of 
gyration. \cite{harn:99} 

Finally, we note that the maximum of $D_c$ in the semidilute entangled regime marks the 
onset of glassy dynamics which is discussed in ref \cite{rauc:03}. The friction-controlled 
dynamics in this concentration regime is not captured by eqs \ref{eq14} and \ref{eq15} 
and will be discussed in subsection \ref{glassydynamic}.

\subsection{Coupling of cooperative fluctuations with single polymer chain motion}
In the following we shall discuss the equation of motion which determines the dynamics 
an individual polymer chain. The PS chains are linear chain molecules which are described 
by a chain model for macromolecules. \cite{harn:95,harn:96,harn:98} We consider a continuous,
differentiable space curve ${\bf r}(s,\tau)$, where $s \in[-L/2,L/2]$ is a coordinate along the
macromolecule and ${\bf r}(L/2,\tau)$ is the position vector of the labeled end monomer. The 
Langevin equation of motion including hydrodynamic interaction is given by \cite{harn:96}
\begin{eqnarray} \label{eq22}
3\pi\eta\frac{\partial}{\partial \tau}{\bf r}(s,\tau)&=&\int\limits^{L/2}_{-L/2}
ds'\,\left(3\pi\eta H(s-s')+\delta(s-s')\right) \nonumber
\\&\times&\!\!\!\left(O(s') {\bf r}(s',\tau) + {\bf f}(s',\tau)\right) + {\bf F}(s,\tau),
\end{eqnarray}
with
\begin{eqnarray} \label{eq23}
O(s)&=&3k_BT p\frac{\partial^2}{\partial s^2} - \frac{3k_BT}{4p}\frac{\partial^4}
{\partial s^4}\,.
\end{eqnarray}
Here $1/(2p)$ is the persistence length, $H(s-s')$ is the hydrodynamic interaction 
tensor, and ${\bf f}(s,\tau)$ is the stochastic force. The force ${\bf F}(s,\tau)$ 
describes the influence of intermolecular forces and is discussed below. The numerical 
solution of eq \ref{eq22} allows one to calculate the mean square displacement 
(see eq \ref{eq4}) according to 
\begin{eqnarray}   \label{eq25}
\phi_s(\tau)=\left\langle \left({\bf r}(L/2,\tau)-{\bf r}(L/2,0)\right)^2\right\rangle\,.
\end{eqnarray}
This chain model has been used in the limit \mbox{${\bf F}(s,\tau)=0$} in order to 
describe FCS measurements of DNA molecules in dilute solution. 
\cite{gros:06,petr:06,wink:06} In particular, the model predicts the observed crossover 
from subdiffusive motion \mbox{($B_s(\tau)$ in eq \ref{eq4})} to diffusive motion 
\mbox{($6 D_s \tau$ in eq \ref{eq4})} upon increasing the time $\tau$. 
Moreover, it has been shown that the chain ends are more mobile than the central 
part of the polymer chain for short times. \cite{wink:06} For comparison 
we note that the quantity $\phi_s(\tau)$ contributes to the so called incoherent 
dynamic structure factor which is accessible by quasielastic neutron 
scattering (see ref \cite{harn:97} and references therein).

The key physics determining the dynamics of chain molecules in semidilute entangled solution 
arises from the intermolecular interaction which are taken into account in terms of the 
force ${\bf F}(s,\tau)$ in eq \ref{eq22}. Various expressions for the force ${\bf F}(s,\tau)$ 
have been proposed 
(see, e.g., refs \cite{schw:89a,seme:91,genz:94,schw:97b,guen:99,seme:98,altu:04,pokr:06,pokr:08}).
These earlier theoretical considerations have demonstrated the coupling of cooperative 
fluctuations with single polymer chain motion in the semidilute entangled regime. 
This coupling allows one to measure $D_c$ from the dynamics of individual labeled 
polymer chains with FCS. Hence, it provides the explanation for the finding of a cooperative 
mode in the FCS-experiment. The topological interaction in semidilute entangled polymer 
solutions seriously affects dynamical properties since it imposes constraints on the 
motion of the polymers. When the motion of a single polymer chain is partly hindered by 
the presence of other chains the cooperative diffusion becomes highly correlated and can 
be studied using only a small fraction of labeled molecules. Moreover, the number of molecules 
statistically involved in the correlated dynamics increases considerably upon approaching 
the glass transition concentration.

\begin{figure}[ht!]
\vspace*{0.5cm}
\begin{center}
\includegraphics[width=7cm]{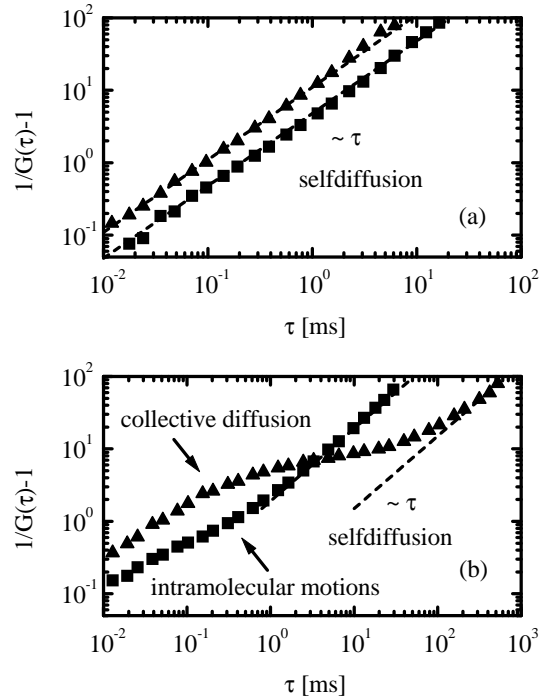}
\caption{The FCS autocorrelation function $1/G(\tau)-1$ of 17 kg/Mol polystyrene 
chains (a) and 515 kg/Mol polystyrene chains (b) in dilute  solution 
($\blacksquare$, \mbox{$c \to 0$}) and in semidilute  solution 
($\blacktriangle$, \mbox{$c = 13 $ wt \%}). The dashed lines of slope $\tau$ 
characterize self-diffusion. Intramolecular motions and cooperative 
diffusion dominate in dilute and semidilute entangled solution, respectively, 
for short times in the case of the high molecular weight polystyrene chains in (b).}
\label{fig8}
\end{center}
\end{figure}

Figures \ref{fig8} (a) and (b) display the function $1/G(\tau)-1$ for the 17 kg/Mol PS 
chains and the 515 kg/Mol PS chains in dilute solution (solid squares, \mbox{$c \to 0$}) 
and in semidilute solution (solid triangles, \mbox{$c = 13 $ wt \%}). For the 
17 kg/Mol PS chains only self-diffusion can be measured using FCS irrespective of 
the concentration (see Figure \ref{fig8} (a)) because of insufficient chain overlap.
In the case of the 515 kg/Mol PS chains self-diffusion dominates for large times as is 
indicated by the dashed lines in Figure \ref{fig8} (b). The cooperative diffusion observed 
in the semidilute entangled solution (solid triangles in Figure \ref{fig8} (b)) dominates 
the autocorrelation function on the same time scale as intramolecular chain relaxations 
in the case of a dilute solution (solid squares in Figure \ref{fig8} (b)). Hence one may 
conclude that upon increasing the polymer concentration the contribution of internal
chain motions to the single chain dynamics decreases while the contribution of the 
cooperative motions increases because of the fluctuations of the surrounding polymer 
chains. Both types of dynamics are observable on the same time scale but in different 
concentration regimes for high molecular weight PS chains. In the case of internal chain 
motions the dynamics is driven by fluctuations of the solvent while fluctuations 
of the surrounding polymer network induce the cooperative dynamics. The fact that 
cooperative diffusion and internal chain motions occur on similar time and length 
scales has already been discussed earlier (see ref \cite{jian:96} and references 
therein).
%
%
%
%

\begin{figure}[t]
\vspace*{0.5cm}
\begin{center}
\includegraphics[width=7cm]{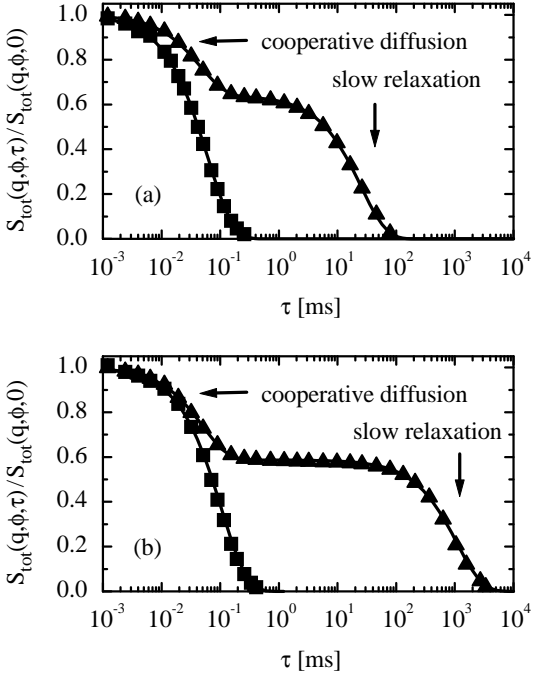}
\caption{The  total dynamic scattering function $S_{tot}(q,\phi,\tau)$ of 
67 kg/Mol polystyrene chains (a) and 515 kg/Mol polystyrene chains (b) measured by 
DLS in semidilute unentangled solution ($\blacksquare$, \mbox{$c = 1 $ wt \%}) 
and in semidilute entangled solution ($\blacktriangle$, \mbox{$c = 13 $ wt \%}). 
The solid lines follow from eq \ref{eq26}. For short times cooperative diffusion 
dominates, while the slow relaxation dominates for very large times in 
semidilute entangled solution. There is no slow relaxation in semidilute unentangled 
solution, i.e., $S_{sl}(q,\phi)=0$ in eq \ref{eq26}. The absolute value of the 
scattering vector is given by \mbox{$q=157.6\, \mu$m$^{-1}$}.}
\label{fig9}
\end{center}
\end{figure}

Without entanglements the local concentration fluctuations at low scattering 
vectors ${\bf q}$ are suppressed by the osmotic pressure of the solution, and the 
total dynamic scattering function $S_{tot}(q,\phi,\tau)$ measured by DLS decays via 
cooperative diffusion according to eqs \ref{eq14} - \ref{eq16}. However, in the presence 
of entanglements, there is an additional suppression of concentration fluctuations.
Some concentration fluctuations may be frozen in by the entanglements.
\cite{broc:77,doi:92,eina:99} This fraction of light scattering signal may only decay
 with the spectrum of relaxation times of the entanglements themselves, leading to a 
slow relaxation of the total dynamic scattering function as is shown in Figures \ref{fig9} 
(a) and (b) for the 67 kg/Mol and 515 kg/Mol PS chains in semidilute entangled solution 
at \mbox{c=13 \% wt} (solid triangles). The corresponding upper solid lines in Figures 
\ref{fig9} (a) and (b) have been calculated according to
\begin{eqnarray}   \label{eq26}
S_{tot}(q,\phi,\tau)&=&S_{c}(q,\phi)\exp(-q^2 D_c \tau)\nonumber
\\&+&S_{sl}(q,\phi)\exp(-\tau/\tau_{sl})\,,
\end{eqnarray}
where $\tau_{sl}$ is a decay time. For arbitrary values of the magnitude of the 
scattering vector $q$ and the volume fraction $\phi$, the shape of the total dynamic 
scattering function $S_{tot}(q,\phi,\tau)$ is more complex than the expression 
given in eq \ref{eq26}. For large values of $q$ intramolecular motions lead 
to a stretched exponential decay of $S_{tot}(q,\phi,\tau)$ for short times 
(see e.g., refs \cite{harn:96,harn:99}). Moreover, the contribution 
of the slow relaxation to $S_{tot}(q,\phi,\tau)$ is in general given by 
a linear combination of exponentially decaying functions, i.e., 
$\sum_i \exp(-\tau/\tau_{i,d})$. \cite{eina:02,take:07} 
%
%
%
%

Experiments on PS in various solvents have confirmed that the slow relaxation
can be measured using DLS.
\cite{brow:90a,nico:90a,nico:90b,nico:90c,wang:93,brow:93,sun:94,wang:95,lin:97}
However, the microscopic understanding of the slow relaxation needs to be improved. 
\cite{li:08} On the basis of our FCS 
and DLS measurements shown in Figures \ref{fig8} and \ref{fig9} we note that 
self-diffusion ($D_s$) occurs on an intermediate time scale, i.e.,
\mbox{$1/(q^2D_c)=0.05$ ms},  \mbox{$1/(q^2D_s)=16$ ms}, and  \mbox{$\tau_{sl}=1087$ ms} 
for $q=157.6 \mu$m$^{-1}$ for the 515 kg/Mol PS chains. For comparison Figures \ref{fig9} 
(a) and (b) also display the measured total dynamic scattering function of the 
PS chains in semidilute unentangled solution (solid squares). In this case there 
is no slow relaxation due to insufficient chain overlap. The corresponding 
lower solid lines in Figures \ref{fig9} (a) and (b)  have been calculated according 
to eq \ref{eq26} with $S_{sl}(q,\phi)=0$.

\begin{figure}[h!]
\vspace*{0.5cm}
\begin{center}
\includegraphics[width=8cm]{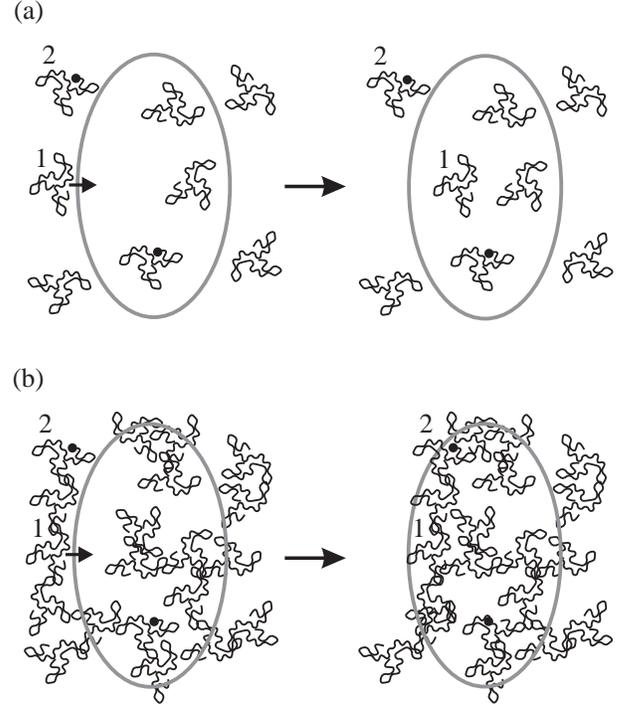}
\caption{Schematic illustration of the cooperative diffusion process which is related 
to the relaxation of the total polymer number density towards the average total number 
density. The polymer chain denoted by the index 1 moves in (a) and (b) from left to 
right into the FCS observation volume enclosed by the grey ellipsoidal lines. The 
polymer chain diffuses into the observation volume nearly without influencing the 
locations of the remaining polymer chains in an unentangled solution in (a), while 
the motion of the polymer chain leads to coherent movement of the surrounding polymer 
chains in semidilute entangled solution in (b). The size of the polymer chains, the 
size observation volume, and the number of polymer chains are not drawn to absolute 
scale. Only the fact that in (b) the motion of the unlabeled polymer chain denoted by 
the index 1 induces a correlated movement of the labeled polymer chain denoted by the 
index 2 into the observation volume is relevant. Each labeled polymer chain carries 
only one dye molecule at one of its ends which is marked by a black dot. As the 
labeled polymer chain denoted by the index 2 diffuses into the observation volume 
from left to right in (b), it causes temporal fluctuations of the detected fluorescence 
intensity which can be measured by FCS even in the case that the number of labeled 
polymer chains is considerably smaller than the number of unlabeled polymer chains.
In addition self-diffusion can be measured using FCS both in (a) and (b) as 
discussed in Section IV A. In (b) self-diffusion of polymer chains corresponds 
to movements of the polymer chains along their contour through the transient 
network.}
\label{fig12}
\end{center}
\end{figure}

The direct DLS measurement of the slow relaxation confirms our earlier remark that 
cooperative diffusion becomes highly correlated in the transient entanglement network and 
can be studied using only a small fraction of labeled polymer chains within FCS. 
As is illustrated in Figure \ref{fig12} (b) unlabeled polymer chains (see, e.g., the polymer
chain denoted by the index 1) and labeled polymer chains (see, e.g., the polymer chain 
denoted by the index 2) move in a coherent manner due to entanglements into the FCS 
observation volume enclosed by the grey ellipsoidal lines. The resulting temporal 
fluctuations of fluorescence light emitted by labeled polymer chains can be detected 
by FCS in terms of the cooperative diffusion. A spherical volume of mean size equivalent 
to the radius of gyration of an individual polymer chain contains about 15 polymer chains 
at the concentration $c^+$ at which cooperative diffusion is measured with FCS. Consequently, 
neighbouring chains strongly interpenetrate and entangle with each other leading to 
highly cooperative motions in this correlated state. Without entanglements 
cooperative diffusion cannot be detected if only a small fraction of the polymer chains 
are labeled due to insufficient chain overlap. Hence in dilute and semidilute unentangled
solutions the unlabeled polymer chain denoted by 1 in Figure \ref{fig12} (a) moves from 
left to right into the FCS observation volume nearly without influencing the remaining 
labeled and unlabeled polymer chains.
%
%
%

\subsection{Onset of glassy dynamics} \label{glassydynamic}
\begin{figure}[h!]
\vspace*{0.5cm}
\begin{center}
\includegraphics[clip,width=7cm]{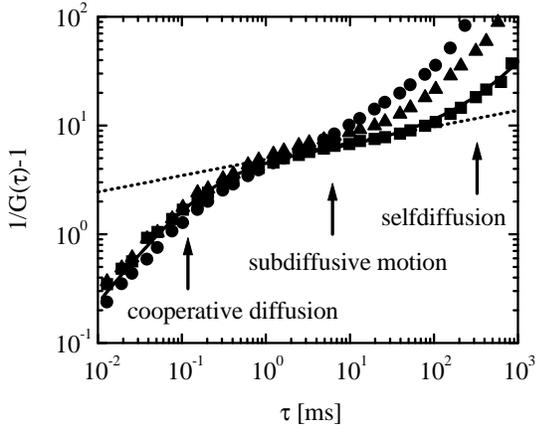}
\caption{The measured FCS autocorrelation function $1/G(\tau)-1$ of a 515 kg/Mol 
polystyrene solution at three concentrations: \mbox{$c = 9.1 $ wt \%}, ($\bullet$);
\mbox{$c = 13 $ wt \%}, ($\blacktriangle$); \mbox{$c = 20 $ wt \%}, ($\blacksquare$).
The solid line displays the result for the highest concentration as obtained from 
eq \ref{eq5} with eqs \ref{eq4} and \ref{eq27} as input. For short and large times 
cooperative diffusion and self-diffusion dominate, respectively. The dotted line 
represents the asymptotic shape of the autocorrelation function in the 
intermediate time regime.}
\label{fig10}
\end{center}
\end{figure}

Upon approaching the glass transition concentration \mbox{$c_{gl}\approx 80$ wt \%}
of PS in toluene, \cite{rauc:03,kona:93} the dynamics of the polymer chains slows down 
considerably (see ref \cite{pete:09} and references therein). A first signature of 
this slowing down is given by the deviations of the measured cooperative diffusion 
coefficients $D_c$ from the solid line at high concentrations in fig \ref{fig7}. 
The cooperative diffusion coefficient decreases by more than three decades as compared 
to its maximum value upon further increasing the concentration  (see fig 6 in ref
\cite{rauc:03}). A second signature of the onset of glassy dynamics is given by the 
shape of the autocorrelation function $G(\tau)$ measured with FCS. Figure \ref{fig10} 
displays measured functions $1/G(\tau)-1$ (solid symbols) for the 515 kg/Mol PS chains 
at three concentrations  \mbox{$c = 9.1, 13$, and $20 $ wt \%} together with the 
autocorrelation function for the highest concentration (solid line) calculated according 
to eq \ref{eq5} with eq \ref{eq3} and 
\begin{eqnarray}  \label{eq27}
\phi_s(\tau)&=&6 D_s \tau+ A_s \tau^\beta\,,\hspace{0.5cm} \beta=0.3\,. 
\end{eqnarray} 
Subdiffusive motion characterized by the stretching parameter $\beta$ is observed 
as an additional mode on an intermediate time scale between the fast cooperative 
diffusion ($D_c$) and the slow self-diffusion ($D_s$). The dotted line in fig \ref{fig10} 
represents the asymptotic shape of $1/G(\tau)-1$ in the intermediate time regime. 
Both the exponent $\beta=0.3$ and the time scale agree with literature values for 
PS. \cite{rauc:03,lind:79}
%
%
%
%

\section{An application: Comparison with minimum concentration required to produce nanofibers}
The understanding of dynamical properties of semidilute entangled polymer solutions 
is also important for various technological relevant applications. As an example 
we discuss the formation of nanofibers from polymer solutions. Polymer nanofibers 
are attractive building blocks for functional nanoscale devices. They are promising 
candidates for various applications, including filtration, protective clothing, 
polymer batteries, sensors, and tissue engineering. \cite{rama:05,stev:05}
Electrospinning is one of the most established fiber fabrication methods and has attracted 
much attention due to the ease by which nanofibers can be produced from polymer 
solutions. \cite{grei:07} Fibers produced by this approach are at least one or two 
orders of magnitude smaller in diameter than those produced by conventional fiber 
production methods like melt or solution spinning. In a typical electrospinning 
process a jet is ejected from the surface of a charged polymer solution when the 
applied electric field strength overcomes the surface tension. The ejected jet travels 
rapidly to the collector target located at some distance from the charged polymer 
solution under the influence of the electric field and becomes collected in the form 
of a solid polymer nanofiber. However, this method requires a dc voltage in the 
kV range and high fiber production rates are difficult to achieve because only a 
single fiber emerges from the nozzle of the pipet holding the polymer solution. 
\cite{grei:07} In order to overcome these deficiencies an efficient procedure 
enabling the parallel fabrication of a multitude of polymer fibers with regular 
morphology and diameters as small as 25 nm has been reported recently. \cite{weit:08}
It involves the application of drops of a polymer solution onto a standard spin 
coater, followed by fast rotation of the chuck, without the need of a mechanical 
constriction. The fiber formation relies upon the instability of the spin-coated 
liquid film that arises due to a competition of the centrifugal force and the 
Laplace force induced by the surface curvature. This Rayleigh-Taylor instability 
triggers the formation of thin liquid jets emerging from the outward driven 
polymer solution, yielding solid nanofibers after evaporation of the solvent.
%
%
\begin{figure}[h!]
\begin{center}
\includegraphics[clip,width=7cm]{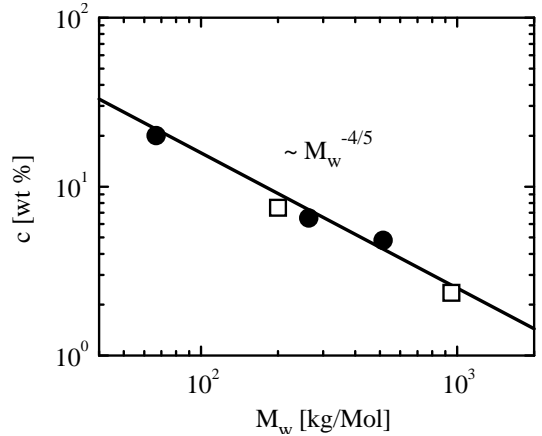}
\caption{The concentration $c^+$ ($\bullet$) at which the cooperative diffusion 
mode appears in the FCS measurements together with the minimum concentration 
$c_{fib}$ ($\square$) required to produce nanofibers \cite{weit:08} as a function 
of the molecular weight $M_w$. The solid line of slope $M_w^{-4/5}$ represents a 
scaling relation valid for polymers in a good solvent.}
\label{fig11}
\end{center}
\end{figure}
%
%

The reason why the ejected jets of polymer solution do not further break up into 
individual droplets, but rather give rise to continuous, solid nanofibers, is the 
related to the dynamic properties of the polymer solutions. In order elucidate this 
point in more detail, Figure \ref{fig11} displays the minimum concentration $c_{fib}$
required to produce nanofibers from 200 kg/Mol and 950 kg/Mol poly-(methylmethacrylate) 
solution (open squares) \cite{weit:08} together with the concentration $c^+$ at 
which the cooperative diffusion mode appears in the FCS measurements of the 67 kg/Mol, 
264 kg/Mol, and 515 kg/Mol PS solutions (solid circles). Interestingly, the 
concentrations $c_{fib}$ and $c^+$ follow approximately the same scaling relationship 
$c_{fib}=c^+ \sim  M_w^{-4/5}$ (c.f., eq \ref{eq8}). Hence, the nanofiber 
formation requires that the polymer concentration exceeds the concentration $c^+$ 
where basically all molecules are involved in the correlated cooperative dynamics. 
Uniform fibers cannot be obtained for lower concentrations due to insufficient 
chain overlap and the dominating self-diffusion which leads to a disentanglement 
under the influence of external forces such as the centrifugal force or the 
electrostatic force.

\section{Conclusion}
A general analysis of the diffusion in polystyrene solutions obtained by 
fluorescence correlation spectroscopy and by dynamic light scattering has been 
presented. Two different diffusion coefficients have been obtained with fluorescence 
correlation spectroscopy using single-labeled polystyrene in toluene solutions 
[Figures \ref{AC} - \ref{fig4}]. The self-diffusion coefficient $D_s$ results from 
fluorescence correlation spectroscopy in the limit of small concentrations of 
labeled molecules and for arbitrary concentrations of unlabeled molecules. Moreover, 
the cooperative diffusion coefficient $D_c$ in the semidilute entangled regime 
becomes accessible as well which is ascribed to an {\bf effective} long-range interaction 
of the labeled chains in the transient entanglement network. The self-diffusion coefficients 
$D_s$  can be determined from the cooperative diffusion coefficient $D_c$ obtained 
by dynamic light scattering measurements and vice versa according to eqs \ref{eq6}
and \ref{eq7}.

The measurements verify the basic scaling and reptation theory for semidilute 
entangled polymer solutions [Figures \ref{DSversusClargeMW}, \ref{fig5}, 
\ref{fig6} and eqs \ref{eq9}, \ref{eq11}, \ref{eq12}]. A quantitative basis for 
the modelling of the cooperative diffusion coefficient is given by a Langevin and 
generalized Ornstein-Zernike equation [eqs \ref{eq13} - \ref{eq21}]. The calculated 
cooperative diffusion coefficients agree with the measured results both in the dilute 
and semidilute regimes [Figure \ref{fig7}]. In particular the features of the 
crossover region between the dilute and the semidilute regimes are captured correctly 
by the underlying integral equation theory. 

For large times the decay of the fluorescence correlation spectroscopy autocorrelation 
function is dominated by self-diffusion, while intramolecular chain relaxations in 
dilute solution and cooperative diffusion in semidilute entangled solution dominate for 
short times [Figures \ref{fig6} and \ref{fig8}]. An additional slow relaxation in 
semidilute entangled solution can be observed by dynamic light scattering [Figure \ref{fig9}]. 
Moreover, the fluorescence correlation spectroscopy autocorrelation function 
exhibits an additional mode on an intermediate time scale upon approaching the glass 
transition concentration [Figure \ref{fig10}].

Finally, it has been shown the minimum concentration required to produce solid 
nanofibers from a polymer solution follows the same scaling relationship 
as the concentration at which the cooperative diffusion mode appears in the fluorescence 
correlation spectroscopy measurements [Figure \ref{fig11}]. The nanofiber 
formation requires that the polymer concentration exceeds the concentration where 
basically all molecules are involved in the correlated cooperative dynamics. 
Hence fluorescence correlation spectroscopy is helpful for the understanding of 
dynamical properties of semidilute entangled polymer solutions in the case of 
technological relevant applications.
\\

We thank A.\ H.\ E.\ M\"uller and A.\ B\"oker for the synthesis of the polymers and the Deutsche Forschungsgemeinschaft, SFB 481 (A11), Bayreuth, for financial support.

\end{document}